\documentclass{article}
\usepackage{arxiv}

\usepackage[utf8]{inputenc} % allow utf-8 input
\usepackage[T1]{fontenc}    % use 8-bit T1 fonts
\usepackage{hyperref}       % hyperlinks
\usepackage{url}            % simple URL typesetting
\usepackage{booktabs}       % professional-quality tables
\usepackage{amsfonts}       % blackboard math symbols
\usepackage{nicefrac}       % compact symbols for 1/2, etc.
\usepackage{microtype}      % microtypography
\usepackage{lipsum}		% Can be removed after putting your text content
\usepackage{graphicx}
\usepackage{doi}
\usepackage{subcaption}
\usepackage[numbers]{natbib}
\usepackage{xcolor,colortbl}

\title{All Polarized but Still Different: a Multi-factorial Metric to Discriminate between Polarization Behaviors on Social Media}

\author{ \href{https://orcid.org/0000-0003-1634-8856}{Célina TREUILLIER}\\
	Université de Lorraine, CNRS, LORIA\\
	Nancy, FRANCE\\
	\texttt{celina.treuillier@loria.fr} \\
	\And
	\href{https://orcid.org/0000-0003-4252-4526}{Sylvain CASTAGNOS} \\
	Université de Lorraine, CNRS, LORIA\\
	Nancy, FRANCE\\
	\texttt{sylvain.castagnos@loria.fr} \\
 	\And
	\href{https://orcid.org/0000-0002-9876-6906}{Armelle BRUN} \\
	Université de Lorraine, CNRS, LORIA\\
	Nancy, FRANCE\\
	\texttt{armelle.brun@loria.fr} \\
}

\renewcommand{\headeright}{}
\renewcommand{\undertitle}{}

\newcommand{\modelName}{GRAIL~}
\newcommand{\modelNameNospace}{GRAIL}

\begin{document}
\maketitle

\begin{abstract}
Online polarization has attracted the attention of researchers for many years. Its effects on society are a cause for concern, and the design of personalized depolarization strategies appears to be a key solution. Such strategies should rely on a fine and accurate measurement, and a clear understanding of polarization behaviors. However, the literature still lacks ways to characterize them finely. We propose \modelNameNospace, the first individual polarization metric, relying on multiple factors. \modelName assesses these factors through entropy and is based on an adaptable Generalized Additive Model. We evaluate the proposed metric on a Twitter dataset related to the highly controversial debate about the COVID-19 vaccine. Experiments confirm the ability of \modelName to discriminate between polarization behaviors. To go further, we provide a finer characterization and explanation of the identified behaviors through an innovative evaluation framework.
\end{abstract}

\section{Introduction}
\label{intro}

Technology has an increasing impact on our everyday life. Its role, associated with the democratization of social media, is more and more questioned. Online media gradually became a persuasive technology through which users can be manipulated, hence undermining social harmony \cite{benkler2018network}. Digital citizenship~\cite{choi2016concept} promotes education on the use of technologies so that citizens are aware of and able to deal with potential threats. Among them, polarization appears as a key societal issue \cite{tucker2018social}. In this work, we refer to polarization as political divergence, or more globally, a division into several sharply contrasting groups of opinions. Polarization is widely studied by researchers in various fields, agreeing on the possible harmful effects on society, and the need to limit it in certain cases~\cite{kubin2021role}. 

Social conflicts sometimes lead to debate and constructive change, provided that everyone is free to express their opinions~\cite{stray2021designingrs}. However, under certain circumstances, it can lead to a bimodal distribution of opinions and ideological extremes \cite{prior2013media}.

To limit user polarization, common depolarizing processes bring global diversity into the recommended content~\cite{helberger2018diversity, lunardi2020filter, heitz2022benefits} to expose users to a broader range of opinions. However, the need for personalized depolarizing strategies has recently been highlighted in literature~\cite{bernstein2020diversity, stray2021designingrs, treuillier2022being}. The literature does offer many polarization metrics, assessing the global polarization of a group of users, a specific topic, or within a whole network \cite{garimella2018quantifying, guerra2013measure, morales2015measuring}. Besides,  only few individual metrics have been proposed ~\cite{becatti2019extracting, cicchini2022news}. However, these existing metrics, both global and individual, only rely on a single factor and struggle to go beyond a simple binary paradigm. For example, some studies focus on the diversity of communities with which users interact \cite{garimella2018quantifying,becatti2019extracting}, while others study the diversity of sources \cite{cicchini2022news}, but they never combine these multiple factors to their full extent. Paradoxically, the literature is fairly consistent in pointing out that multiple factors, both human and algorithmic,  drive polarization~\cite{geschke2019triple, jost2022cognitive, valensise2023drivers}. In our view, the implementation of personalized depolarization strategies should thus rely on a fine-grained and multi-factorial assessment of polarization. A recent study carried out in a social media context, strengthened the interest of a multi-factorial approach in modeling and understanding polarization behaviors~\cite{phillips2023organizational}. 

Based on these findings, we advocate the need for a new polarization metric, going beyond the existing ones by assessing the polarization of each individual independently and taking into account the multiple factors that influence this polarization. In that respect, we raise two research questions: \textbf{(RQ1) How to combine multiple factors to form a discriminative individual polarization metric?} and \textbf{(RQ2) Does the proposed metric allow to explain polarization behaviors?} 

We propose the \modelName (GeneRalized  AddItive poLarization) metric.  \modelName is evaluated on a social media dataset related to the COVID-19 vaccine debate on Twitter ~\cite{treuillier23}, and its ability to be explained by polarization-specific behavioral indicators for different classes of behavior is assessed. The contributions of this work are thus as follows: {\bf (1)} the definition of \modelNameNospace, the first individual and multi-factorial polarization metric, and {\bf (2)} an innovative evaluation framework allowing the explanation of behavioral classes based on individual polarization scores computed with \modelNameNospace. %The extent to which \modelName is generalizable to other datasets and contexts is further discussed. 

The remainder of this paper is organized as follows. In Section~\ref{SOA}, we provide a literature review of polarization assessment. We then detail the \modelName metric in Section~\ref{metric}. The evaluated data is presented in Section~\ref{data}, and \modelName is evaluated in Section~\ref{caseStudy}. Finally, conclusions and perspectives are drawn in Section~\ref{CCL}.

\section{Literature Review} 
\label{SOA}

\subsection{Online Media and Polarization}

Operating in many contexts (online/offline, individual/group), polarization has raised the attention of a large variety of fields, \textit{e.g.} mathematicians, physicians, psychologists, computer scientists, \ldots. Among other things, many mathematical models of opinion dynamics and polarization have been described for years~\cite{sirbu2017opinion}. Physicians such as Baumann~\textit{et al.}~\cite{baumann2020modeling} show that three scenarios occur in a social media context: consensus, unilateral polarization, or bilateral polarization. Besides, psychologists, e.g. Geschke~\textit{et al.}, propose a three-dimension modeling of polarization: individual, social and technological~\cite{geschke2019triple}.
The technological dimension is reinforced by the daily use of online media, which has considerably changed the way people access the latest news and get information. Two main types of online content broadcasters can be distinguished: online media and social media.

First, online media and news aggregators offer users the opportunity to access a large amount of information every day. This led to the emergence of News Recommender Systems (NRS), which assist users in finding news of interest. However, NRS have also fostered individual polarization~\cite{chen2019modeling, garimella2021political} by creating filter bubbles~\cite{michiels2022filter, pariser2011filter}. The role of diversity in NRS regarding polarization decrease is highly discussed and even highlights a potential strengthening of polarisation \cite{bail2018opposing}. Besides, Wu~\textit{et al.} claim that the diversification process should be personalized for each user according to her behavior~\cite{wu2018personalizing}. In our view, this highlights the need for a comprehensive and individual modeling of polarization.

Second, social media, which are the focus of this work, are now commonly used by a large proportion of the population and are undoubtedly involved in the polarization process~\cite{kubin2021role}. The following section is dedicated to social media.

\subsection{Polarization Metrics on Social Media}

Social media have been democratized since the late 2000s. They are now used on a daily basis all around the world. Although social media have strong benefits such as liberty of interactions, free access to information and democratization, they also play a critical role in the polarization process~\cite{kubin2021role,vanbavel2021media}. From the very first steps of Twitter, Conover~\textit{et al.} studied polarization on social media~\cite{conover2011political}. They have shown that public political interaction (mentions and retweets) induces distinct network topologies. The {\it retweets} network allows to distinguish two well-separated communities, thus showing that there is a high polarization, whereas the {\it mentions} network results in a unique highly connected community. The authors rely on modularity to quantify polarization~\cite{newman2006modularity}, thus capturing the strength of the division of a network into modules (communities). While giving some insights about the division of a network, Guerra~\textit{et al.} explain that modularity does not seem to be a direct measure of polarization, as non-polarized networks can also be divided into distinct communities~\cite{guerra2013measure}. The authors also emphasize the potential role of community boundaries to quantify polarization, by comparing internal and external connections of intermediary nodes. Polarization can also be evaluated through the {\it controversy} of a topic~\cite{garimella2018quantifying}, \textit{i.e.} its propensity to provoke heated debates. Highly controversial topics automatically lead to well-separated communities in the network, and \textit{controversy} can thus be measured, for example, by examining the information flow between communities. Morales~\textit{et al.} propose to model opinions shared in a social media by the probability density function:  polarization is then quantified according to the resulting distribution~\cite{morales2015measuring}. The authors also propose to consider attractors, \textit{i.e.} a concentration of users around different opinion positions, in polarization metrics~\cite{morales2015measuring}. 
We can conclude that polarization metrics from the social media literature rely on a varied set of information. Though, these measures share one major characteristic: they all quantify polarization or controversiality of a whole network or topic. In the form they are presented, existing polarization metrics thus provide an overview of the state of polarization in a network or about a specific topic, but give no information about the level of polarization of each user, individually.

Recall that polarization is influenced by both individual and group factors \cite{geschke2019triple}. We thus promote individual and group polarization metrics to participate in a refined model of polarization. However, such metrics remain uncommon in the literature. Among the few existing ones, we can find the \textit{polarization score} proposed by Becatti~\textit{et al.}~\cite{becatti2019extracting}. It is based on the differentiation of a set of communities $C$ from the graph built from users' interactions. To evaluate this metric, a user $u$ is represented by a vector $I_{u,C}$ where each element $I_{u,c}$ represents the proportion of interactions of $u$ with community $c$ ($N_{u,c}$), compared to her total number of interactions $N_u$. The polarization score $\rho(u)$ of user $u$, is evaluated as the maximum value of vector $I_{u,C}$, as presented in Equation~(\ref{becatti_metric}).

\begin{equation}
\label{becatti_metric}
    \rho(u)= \max_{c \in C}\big\{\frac{N_{u,c}}{N_u}\big\}
\end{equation}

 Users with $\rho \approx 1$ are thus users who access a unique community and are highly polarized. Users who access all the communities equally, have $\rho = 1/C$. In a two-community context, the polarization score presented by Schmidt~\textit{et al.} is similar but is oriented, in the sense that the value informs which community is accessed the most~\cite{schmidt2018polarization}.

Cicchini~\textit{et al.}~\cite{cicchini2022news} proposed the {\it Lack of Diversity} (LD) metric. It is highly similar to the \textit{polarization score} of Becatti~\textit{et al.}~\cite{becatti2019extracting}, but considers sources of information instead of communities while correcting the bias linked to the popularity of these sources. In their work, sources of information consist in a set of $M$ media outlets, with which users can interact. Thus, each user $u$ is represented by her number of interactions $N_{u,m}$ with news from media $m$. $LD$ is computed as follows: 
\begin{equation}
\label{cicchini_metric}
    LD(u) = \max_{m \in M}\big\{N_{u,m} \cdot log(\frac{|U|}{|U_m|})\big\} 
\end{equation}

$U$ is the set of users of size $|U|$, while $|U_m|$ is the number of users interacting with media $m$. The $log$ term corrects a potential bias introduced by an imbalance in the number of interactions with a specific source $m$. The LD metric is not bounded.

In our view, the aforementioned individual polarization metrics face some limits. First, both \textit{polarization score}~\cite{becatti2019extracting} and \textit{Lack of Diversity}~\cite{cicchini2022news} only exploit the maximum value in each user vector. It is very reductive: two users who behave the same way in their main community/media will get the same metric value, while they may behave quite differently in other communities/media. This is not taken into account in the metrics, which we believe limits the accuracy of the modeling. Indeed, all values can bring useful insight, and they can all contribute to a better understanding of polarization behavior. Second, these metrics are only measured on a single factor (interaction with communities or interaction with sources), while polarization occurs over the influence of multiple factors~\cite{chen2019modeling, sunstein1999law}. As a consequence, if only a single factor is considered, several users may be identified as similarly polarized according to this factor, but may in fact exhibit a wide range of behaviors and distinguish themselves according to other factors. To sum up, despite a strong community dealing with polarization, the literature still lacks an individual and multi-factorial polarization metric.

\section{\modelNameNospace: A New Individual and Multi-Factorial Polarization Metric}
\label{metric}

In this work, we propose \modelName (\textit{GeneRalized AddItive poLarization}), a new polarization metric designed to finely measure the degree and nature of polarization of each user. The degree corresponds to the intensity of the polarization, while the nature informs about the community a user is closer or belongs to. 
 \modelName has several features: {\bf (F1)} multi-factorial, \textit{i.e.} it can consider multiple polarization factors,  {\bf (F2)} discriminative, \textit{i.e.} it goes beyond a polarized/non-polarized distinction and allows a fine modeling of polarization, and {\bf (F3)} generalizable, \textit{i.e.} it can be used on any type of data and data source.

The following subsections detail the basic components of \modelNameNospace, that rely on 1) an information theory-related way to manage factors of any type and foster fine-grained modeling of user behavior, 2)  a polynomial transformation to heighten the discrimination between users, 3) a Generalized Additive Model to combine multiple factors. To guide the author through this section and illustrate the components that constitute the proposed metric, we punctuate our explanations with a running example.

\subsection{An Entropy-Based Metric to Better Match User Behaviors}
\label{entropy}

Let $Z$ be a factor, composed of a set of entities with which users can interact. We propose to represent this factor as a probability distribution to represent the diversity/heterogeneity of a user's interactions with these entities. 

Following Information Theory concepts, this diversity can be evaluated through entropy~\cite{shannon1948mathematical}, a measure of the uncertainty of a data source.  
The more homogeneously distributed the probability mass, the higher the entropy and the greater the uncertainty. Entropy is maximum in the case of the equiprobability of entities. Conversely, if an important part of the probability mass is spread over a small set of entities, the entropy is low. The entropy is 0 if an entity has a 1 probability for one of its entities.
As the maximal entropy value is a function of the number of entities, we use the normalized entropy (see Equation~(\ref{eq:entropy})):

\begin{equation}
\label{eq:entropy}
  H_N(Z)= \frac{-\sum_{z} P(z)log(P(z))}{log(n)}
\end{equation}

Where $Z$ is a discrete random variable that takes $n$ possible entity values, and $P(z)$ is the probability of entity $z$. For the sake of simplicity, we will refer to normalized entropy as entropy. Entropy can be applied to any factor that can be seen as a random variable (topics in news, shared data sources, accounts followed, etc.), which meets characteristic {\it F3}. 

Let us introduce here our running example.  $Z$ is a set of information sources with which users can interact and $n = 4$. According to their interactions with each of these sources, the probability distribution can be computed for each individual user and forms a probability vector of size 4, with $P(z)$ the probability that the user interacts with the source $z$. Table~\ref{tab:example_users} presents an example of such vectors, computed for four users $u_1$ to $u_4$, with associated probability distributions. The resulting scores computed with existing individual polarization metrics $\rho$\cite{becatti2019extracting} and $LD$ \cite{cicchini2022news}, as well as the normalized entropy, defined in Equation~(\ref{eq:entropy}), can thus be computed for each user and are given in Table~\ref{tab:example_users}. To simplify this example, we are assuming that the four sources $z$ are all equally popular. The correcting term of $LD$ (Equation (\ref{cicchini_metric})) is therefore equal to $1$, and $\rho = LD$. 

\begin{table*}[ht]
\centering
\begin{tabular}{|c|c|c|c|c|}
\hline
\textbf{User} & \textbf{Probability distribution} & \textbf{Baselines $\rho$ and $LD$} &\textbf{Entropy value $H_N(Z)$}  & \textbf{Transformed entropy value $f(H_N(Z))$} \\ \hline
\hline
\textit{\textbf{$u_1$}} & \includegraphics[width=0.15\linewidth]{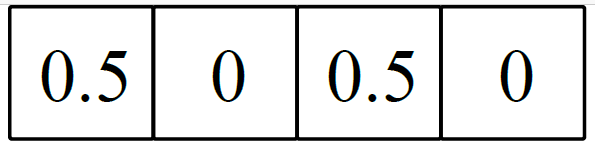} & 0.5 & 0.50 & 0.50 \\ \hline
\textit{\textbf{$u_2$}} & \includegraphics[width=0.15\linewidth]{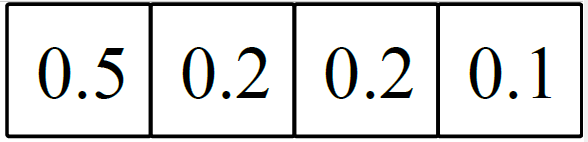} & 0.5 & 0.88 & 0.73 \\ \hline
\textit{\textbf{$u_3$}} & \includegraphics[width=0.15\linewidth]{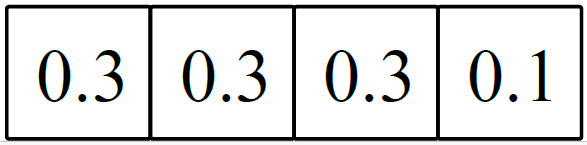} & 0.3 & 0.95 & 0.81 \\ \hline
\textit{\textbf{$u_4$}} & \includegraphics[width=0.15\linewidth]{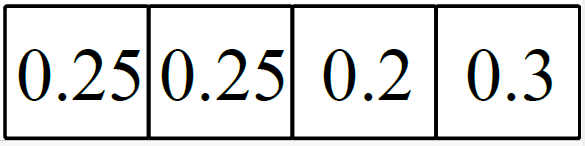} & 0.3 & 0.99 & 0.90 \\ \hline
\end{tabular}
\caption{Example of four users with different behaviors}
\label{tab:example_users}
\end{table*}

Recall that these baseline metrics ($\rho$ and $LD$) only exploit the maximum probability in each vector. The remaining probabilities, whatever their distribution, do not influence the resulting polarization measure. For example, users $u_1$ and $u_2$ have the same maximum value (equals to $0.5$) so both users will be associated with the same metric value, while they interact quite differently on other sources. Regarding entropy, which considers the distribution of all entities of the vector, users $u_1$ and $u_2$ greatly differ. The use of entropy may thus lead to a more accurate metric.

\subsection{A Polynomial Function to Discriminate between Users}
\label{trans_sigmoid}

Recall that one of the objectives of \modelName is to better discriminate between users ({\it F2}). The use of entropy was a first step in that direction. However, part of users may be tightly bunched in the polarization representation space. This limits the differentiation between them. To ensure a wide distribution, we propose to apply a non-linear transformation of the entropy values of the entire set of users to better discriminate between them.

We took inspiration from the logistic function, often used to model non-linear relationships between variables. It represents the cumulative distribution function of the logistic law that is of interest to us. Such a function is bounded, monotonically increasing, has exactly one inflection point at the midpoint, and is differentiable. Its first derivative is a bell shape and its curve has a  S-shaped characteristic. 

\begin{figure}[h!]
\centering
\begin{subfigure}[t]{0.5\textwidth}
  \includegraphics[width=0.85\linewidth]{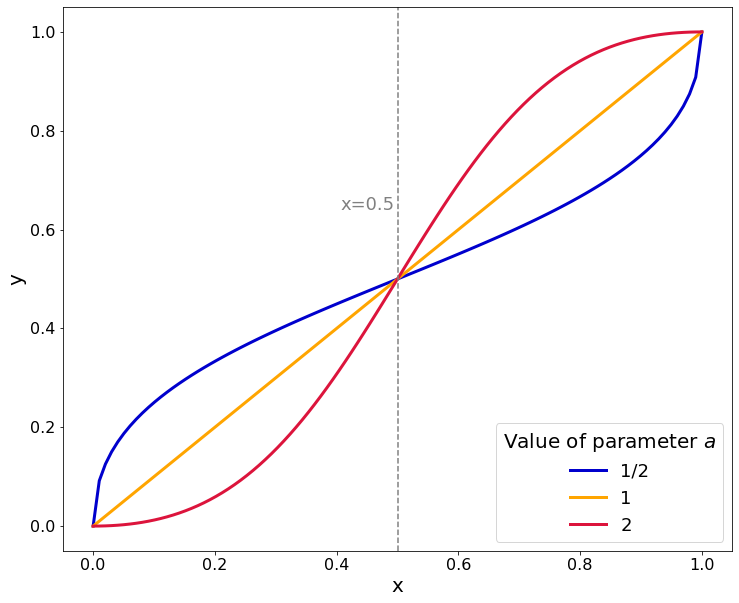}
  \caption{Polynomial curve according to the value of parameter $a$.}
  \label{fig:parameter_a}
\end{subfigure}
\begin{subfigure}[t]{0.5\textwidth}
    \centering
    \includegraphics[width=1\linewidth]{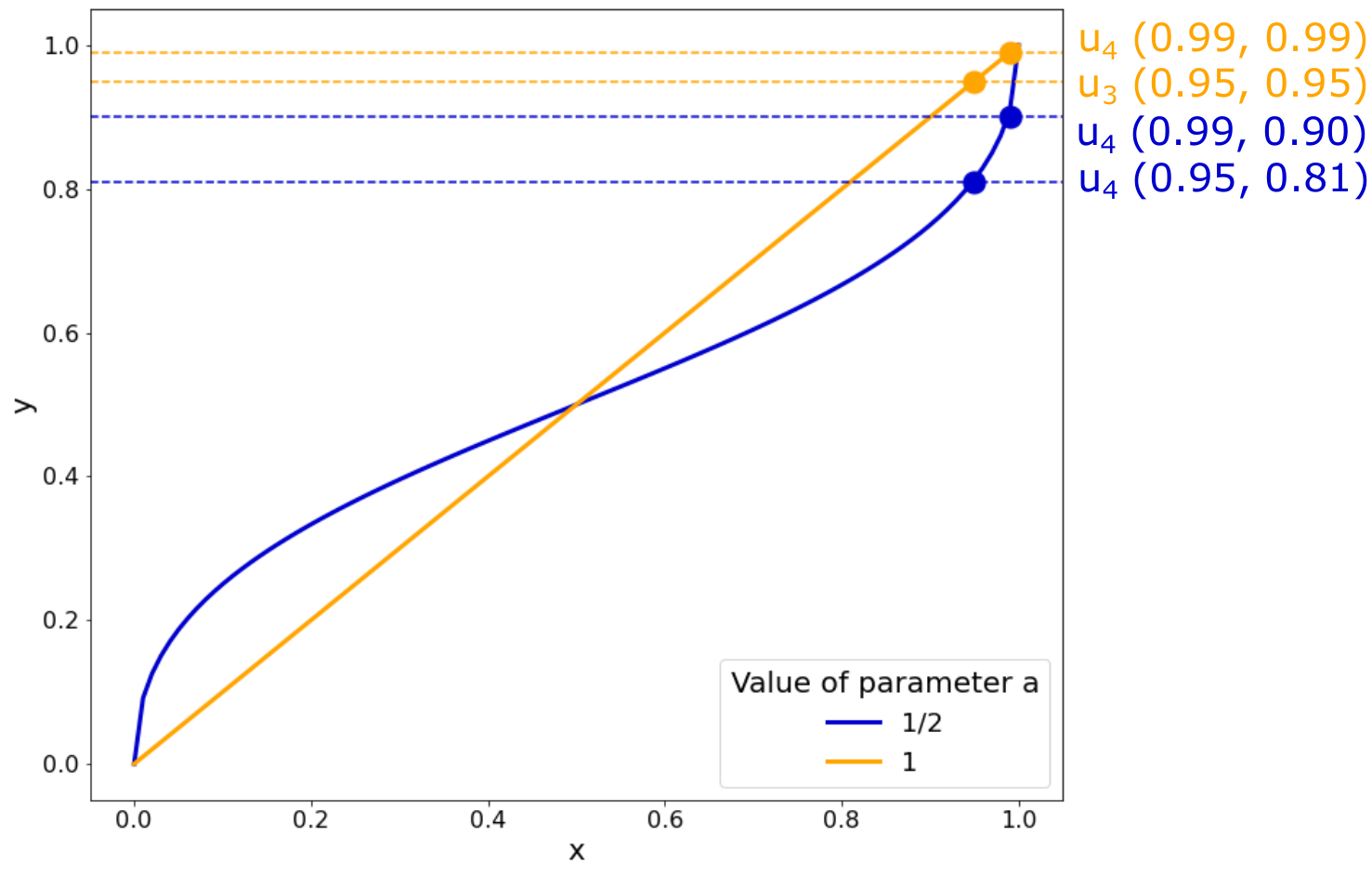}
    \caption{Transformation of entropy values for example users $u_3$ and $u_4$.}
    \label{fig:poly_ex}
\end{subfigure}
\caption{Polynomial function.}
\label{fig:clusters}
\end{figure}

Nevertheless, the standard logistic function has two asymptotes with equations $y=0$ and $y=1$. We intend that \modelName scores range in $[0,1]$, 0 indicating a lack of polarization, and 1 indicating extreme polarization, in line with measures from the literature \cite{becatti2019extracting, cicchini2022news}. We therefore look for a numerically stable function $f$ with $f(0)=0$ and $f(1)=1$, and turn to a function with a similar sigmoid pattern. In accordance with the above-mentioned constraints, we propose to rely on the polynomial function (Equation~(\ref{sigmoid})) to perform the transformation of entropy values. 

\begin{equation}
\label{sigmoid}
    f(x) = \frac{x^a}{x^a + (1-x)^a}
\end{equation}

where $x^a$ controls the yield curve and replaces the traditional $e^x$. In other words, the stiffness of Equation~(\ref{sigmoid}) is described by the parameter $a$. Precisely, with $a=1$, the function is linear. The higher $a$, the steeper the curve and the more sensitive to small changes in $x$ near the curvatures. With $a < 1$ the curve becomes more sensitive to changes in extreme values of $x$ (See Figure~\ref{fig:parameter_a}). 

Let us return to our running example. Equation (\ref{sigmoid}) is applied to the entropy value of each user. First, the polynomial transformation has no impact in $u_1$, whatever the value of parameter $a$, as the entropy of $u_1$ is $0.5$. Besides, given $a=0.5$, the transformed entropy value of $u_2$, $u_3$, and $u_4$ are given in the last column of Table \ref{tab:example_users}. It is worth noting that $u_3$ and $u_4$, the two users with extreme and similar entropy values, are better distinguished once the polynomial function we propose is applied (See Figure \ref{fig:poly_ex}).

In Section~\ref{SOA}, we have highlighted that individual polarization metrics of the literature do not allow for clear discrimination between users. %and often simply indicate whether users are polarized or not %, in an almost binary way. 
By applying the proposed polynomial function (See Equation (\ref{sigmoid})), we aim to better discriminate users with close entropy values ({\it F2}). By mastering the sensitivity of the applied polynomial function thanks to parameter $a$, we can thus adapt this fine discrimination process to a specific set of users.

\subsection{A Generalized Additive Model to Handle Multiple Factors}
\label{GRAIL}

Polarization can occur in multiple contexts and can be influenced by a wide range of factors. We advocate for polarization metrics that consider this multiplicity. These factors are not necessarily related in a linear way, as they can observe unstable variations, and their relation can be particularly difficult to model. For example, the diversity of sources of information a specific user accesses may not be in a linear relationship with the diversity of the topics she is interested in. 

In this respect, Generalized Additive Models (GAMs) appear as a suitable tool to combine multiple dimensions in a unique metric~\cite{hastie2017generalized} (\textit{F1}). GAMs allow the modeling of complex relationships between a response variable $Y$, and several predictor variables $X_i$. The relationship between each variable $X_i$ and $Y$ is modeled using a complex function named a \textit{smooth function} $f_i$. These smooth functions are often represented as splines which are piecewise polynomial functions. These functions can be combined in a unique model, or be added one at a time for each predictor variable. In our work, we model the relationship between factors $X_i$, related to users' behavior, and polarization $Y$. That is why the flexibility offered by GAMs is of particular interest and meets {\it F1} and {\it F3}. 
Taking the structure of GAMs, \modelName is defined as follows:

\begin{equation}
\label{base_metric}
    \modelName(u) = \sum_i \alpha_i f_i(X_{u,i}) + \beta \textrm{ with} \quad \sum_i \alpha_i = 1\
\end{equation}

where $u$ is a user, $f_i$ is a smooth function, $X_i$ is a predictor variable, $\alpha_i$ is the weight of predictor variable $X_i$ and  $\beta$ is a bias term. We use the polynomial function defined in Equation~(\ref{sigmoid}) as a smoothing spline. As we expect the value for a non-polarized user to be 0, we choose to set the bias term  to $\beta=0$. 
Besides, if $\alpha_i=\alpha_j~\forall (i,j)$, it comes down to a traditional GAM. 

To go further, highly polarized topics most often confront two opposing communities: {\it pro} and {\it anti}. For example: pro-life/anti-life, pro-guns/anti-guns, pro-vaccine/anti-vaccine, etc. We thus propose to adapt \modelName so that it not only informs about a user's degree of polarization but also her community of belonging. Thus, \modelName is evaluated as:
\begin{equation}
\label{final_metric}
    \modelName(u) = sgn(p) * \sum_i \alpha_i f_i(X_{u,i})\textrm{ with} \quad \sum_i \alpha_i = 1\
\end{equation}

\begin{equation}
  \textrm{} sgn(p) = \left\{\begin{array}{ll}
         -1& \mbox{if $u$ belongs to the anti community,}\\
         +1& \mbox{if $u$ belongs to the pro community.}
    \end{array}
    \right.
\end{equation}

In the polynomial function $f$ (Equation (\ref{sigmoid})), we propose to replace the variable $x$ by $(1~-H_N~)$, noted $H'(X)$ below, so that higher scores are associated to polarized users. The resulting function is:%we propose to use $1-H_N$ (noted $H'(X)$) so that higher scores are associated to polarized users. The resulting function is:

\begin{equation}
    f_i(X_{u,i}) = \frac{H'(X_{u,i})^{a_i}}{H'(X_{u,i})^{a_i} + (1-H'(X_{u,i}))^{a_i}}
\end{equation}

Altogether, the basic constituents of \modelName make it a good candidate to better discriminate between polarization behaviors. This holds the promise of a better understanding of polarization.

Returning to the running example, additional factors could be computed for users $u_1$ to $u_4$, such as the diversity of topics, communities, sentiment\ldots. As a final step, these multiple factors are combined using the GAM to compete \modelNameNospace. The resulting value is thus multi-factorial (\textit{F1}), and allows a clear distinction between users (\textit{F2}).

We would like to point out that both predictor variables $X_i$ and smooth functions $f$ can be instantiated by any other equation ranging $[0,1]$ to fit any other research context and associated data (\textit{F3}). In a social media analysis context, the factors could for example be graph-related measures (e.g. centrality, PageRank, etc.). 

As a final step, \modelName parameters can be tuned to better fit a given dataset. First, $\alpha$ can be fixed to correspond to research needs or to reflect specific behaviors observed in a dataset. Second,  $a$  (in Equation~(\ref{final_metric})) can also be tuned according to the distribution of data values. Especially, if many users have intermediate values,  $a>1$  will help to better discriminate them, while reassembling users having extreme values. On the contrary, in a context with few intermediate values, $a<1$ should be preferred so has to keep a differentiating power on extreme values. That way, the tuning phase on the $a$ parameter can also serve to evaluate the polarizing power of a topic within the studied dataset. 
The optimization phase for these parameters may be based on observations or on expert knowledge in a controlled environment. Optimally, several objective functions can be defined, and parameter values set according to results.

\section{Data: A Social Media Context}
\label{data}

Social media represent a rich source of data and are commonly studied in the literature to observe polarization behavior. Twitter is one of these social media, where users can share their opinions in concise texts, and where public personalities (politicians, journalists, doctors\ldots) are very active. In our work, we use the dataset presented by Treuillier \textit{et al.}~\cite{treuillier23}. The latter is about the COVID-19 vaccine debate, and data was collected based on the elite users concept\footnote{Popular and legitimate users to address the vaccine debate, being either pro- or anti-vaccine.}~\cite{primario2017measuring}. 

The collected dataset contains 6,697 tweets published during a 7 months period, from January 1st to July 31st, 2022. Among them, 1,869 tweets are from the pro-vaccine elite users, while 4,828 are from anti-vaccine ones. Among all standard users having retweeted these tweets, 1,000 have been studied. The latter have made 16,791 retweets on pro-vaccine tweets, and 283,088 on anti-vaccine tweets. 

As an exploratory step, we evaluate to what extent this dataset actually represents a highly polarized environment. We represent associated data as a graph made of 1,020 nodes (1,000 standard users + 20 elite users), connected by 7,242 edges, corresponding to standard users' retweets on elite users' tweets. There are 2 highly connected sets of nodes (\textit{modularity}=0.55 \cite{clauset2004finding}), and few edges between them. Besides, the controversy of the vaccine debate computed using the Random Walk process described by Garimella~\textit{et al.}~\cite{garimella2018quantifying} is high ($\approx 0.9$), reflecting a minimal number of interactions between the two communities.
These results thus confirm that data are consistent with a highly polarized environment, where anti-vaccine users are much more active than pro-vaccine users, and with few intermediate users, {\it i.e.} users at the interface between the two communities.

\section{Experiments}
\label{caseStudy}

This section is dedicated to the evaluation of \modelName on the previously described dataset. 

We choose to use three relevant factors: {\bf (1)} \textbf{An opinion factor} where opinion represents a community (pro or anti-vaccine). This factor is assessed from the standard users' retweets on each community; {\bf (2)} \textbf{A pro-vaccine source factor}, where elite users who publish information in favor of the COVID-19 vaccine act as pro-vaccine sources. This factor is assessed from the standard users' retweets on each pro-source; {\bf (3)} \textbf{An anti-vaccine source factor}, where elite users who publish against the COVID-19 vaccine act as anti-vaccine sources. This factor is assessed from the standard users' retweets on each anti-source.

\modelName is thus evaluated from these three factors. The opinion factor is computed on the probability distribution over the two communities. The pro-vaccine (resp. anti-vaccine) source factor is computed on the probability distribution over pro-vaccine (resp. anti-vaccine) sources of length 10. 

To evaluate \modelNameNospace, we proceed in three steps. First, we focus on the distribution of the values of each factor. Second, we study the ability of these factors to collectively discriminate between classes of users. Last, we assess the relevance of \modelName in each class of users.

To make this evaluation accurate, we compare \modelName to another three-factor-based metric, that will be referred to as the baseline metric. In this baseline metric, an opinion factor, a pro-vaccine source factor, and an anti-vaccine source factor are also considered, but they are computed by single-factor metric from the literature.  Concretely, $\rho$ is used for the opinion factor, $LD_{pro}$, that corresponds to $LD$ evaluated on pro-vaccine sources, is used for the pro-vaccine source factor and $LD_{anti}$ is used for the anti-vaccine source factor. This comparison will contribute to highlighting the impact of the use of entropy and the transformation with the polynomial function.

\subsection{Distribution of Polarization Factors}
\label{distribution}

\begin{figure}[h!]
    \centering    
    \includegraphics[width=1\linewidth]{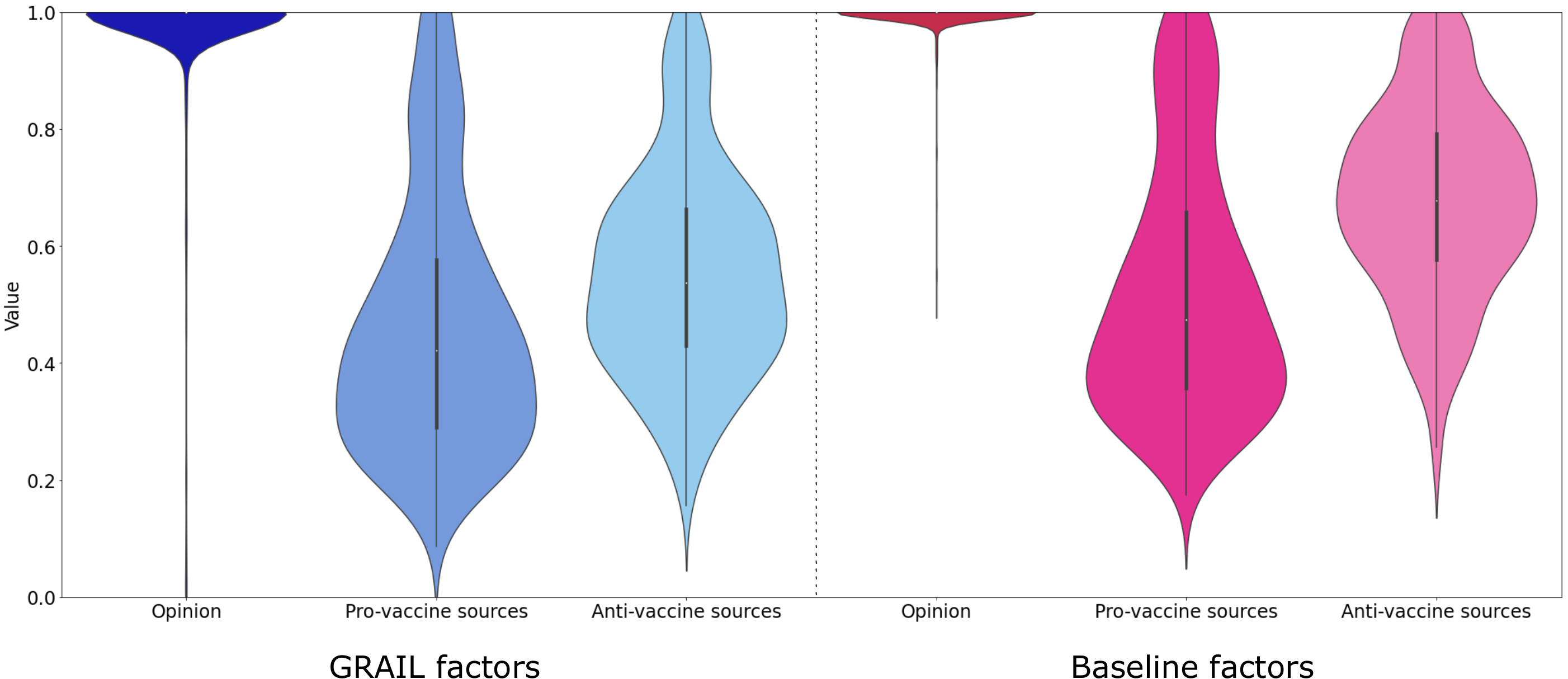}
    \caption{Violin plots of the distribution of polarization factors for \modelName and baseline metrics.}
    \label{fig:distributions}
\end{figure}

In this section, we compare the distribution of factors (opinion, pro-source, and anti-source) computed from entropy-based factors ($H'o$, $H'_{s,pro}$, $H'_{s,anti}$) and from the baseline metric ($\rho$, $LD_{pro}$, $LD_{anti}$).   The distribution of these factors is presented in  Figure~\ref{fig:distributions}.

Let us start by focusing on the opinion factors. We would first like to remind that both \modelName and baseline opinion factors are correlated on a length-2 vector. Indeed, even though the \textit{polarization score} $\rho$ only ranges in $[0.5,1]$, the value of the second element in the vector (used in entropy) can be deduced from the first one, which leads to close distributions. More precisely, the ranking between both factors is the same, even though the associated values are different.
Second, we can see that both distributions are tight.
This reflects the fact that standard users mainly have an activity in a single community and confirms that the COVID-19 vaccine debate is highly polarized. 

Let us now focus on the source factors. About the \modelName source factors, we can see that they are more distributed than the \modelName opinion factor, with the mass of values distributed in the entire range [0;1]. 
In the anti-vaccine source factor, most of the values are around 0.35, whereas they are around 0.55 in the pro-vaccine factor. This means that users tend to retweet a larger number of sources in the pro-vaccine sources than in the anti-vaccine sources (whatever is their respective number of retweets in either community). We can conclude that users tend to be less polarized in the pro-vaccine factor than in the anti-vaccine factor when considering the source factors.
Comparing these \modelName source factors with those of the baseline metric, they are more widely distributed than baseline factors. This is first due to the LD equation, that bounds values in [0.1;1]. Beyond this, in the anti-vaccine factor, \modelName values are significantly more widely distributed than in the baseline anti-vaccine factor. 

To summarize, we confirm that \modelName factors are more widely distributed than baseline factors, which confirms the added value of entropy when we consider to discriminate users. We thus wonder how these factors impact the discrimination of polarization behaviors classes. We propose to compare clusters obtained with the \modelName factors to those obtained with baseline factors.

\subsection{Discrimination of Polarization Behavior Classes}
\label{clustering}

We look for the parameters combination that allows to better discriminate between polarization behaviors. We thus rely on a clustering process, applied to 3D data points: opinion factor, pro-vaccine source factor, and anti-vaccine source factor. We start by tuning \modelName parameters, as explained in Section \ref{GRAIL}.

\subsubsection{Parameters tuning}
\label{parameters_tuning}

Two parameters are optimized: parameter $a$ in Equation~(\ref{sigmoid}), which influences the shape and the slope of the sigmoid-like curve, and therefore the way entropy-based terms are transformed; and parameter $\alpha$, which is used to weight the factors combined in \modelNameNospace.  

To perform the tuning of  $\alpha$, we make it vary between $0$ and $1$ with $\Delta_\alpha = 0.1$. The value $\alpha$ is set on the first factor ($H'_o$), and $(1-\alpha)$ is evenly distributed on the sources terms ($H'_{s,pro}$ and $H'_{s,anti}$). We make $a$  vary between the following values: $[1/4, 1/3, 1/2, 1, 2, 3, 4]$. In total, 77 combinations of parameters are tested.

Both baseline and \modelName metrics are studied.
Note that, to be consistent with the \modelName metric which gives information about the community of belonging of the user (See Section~\ref{GRAIL}), the opinion factors are oriented and normalized in $[0,1]$ for the clustering process. The resulting factor equals 0 if the user is polarized in the anti-vaccine community, and equals 1 if she is polarized in the pro-vaccine community.

Once the three factors, either baseline or \modelNameNospace, are computed for each standard user, we run the traditional $k$-means algorithm~\cite{likas_global_2003} and select the parameters combination that allows maximizing clustering performance, evaluated through well-known Davies-Bouldin~\cite{davies1979cluster} and Silhouette~\cite{rousseeuw_silhouettes_1987} indexes. 

Considering the baseline metric, the optimal value of $\alpha$ is $0.5$, allowing the identification of 2 clusters ($k=2$) with a Silhouette Index~$=0.80$ and a Davies-Bouldin Index~$= 0.32$. Besides, for \modelName factors, optimal performance is reached with $\alpha = 0.6$ and $a = 1/2$. It corresponds to a Silhouette Index~$= 0.85$ and Davies-Bouldin Index~$= 0.35$. These values are obtained with $k=4$ clusters. The optimal value $a=1/2$ confirms that the transformation of entropy-based factors following the polynomial function in Equation~\ref{sigmoid} allows to better discriminate between polarization behavior classes. Besides, as the optimal value of $a < 1$, it confirms that users interact on a highly controversial debate, on which most of them are strongly polarized, while users with intermediate scores are few in number (See Section \ref{trans_sigmoid}).

\subsubsection{Interpretation of Identified Polarization Behavior Classes}

\begin{figure}[h!]
\centering
\begin{subfigure}[t]{0.55\textwidth}
  \includegraphics[width=0.85\linewidth]{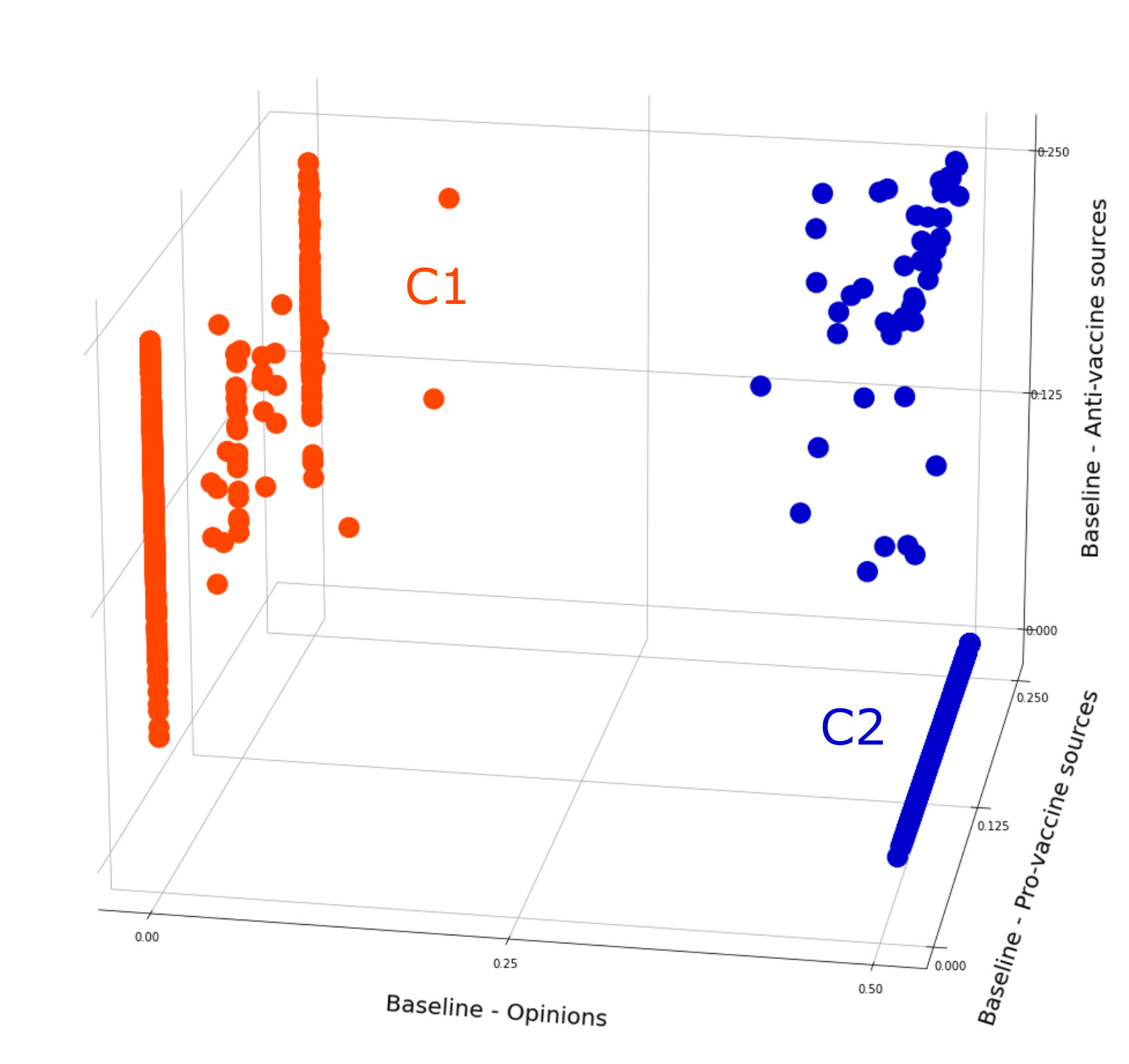}
  \caption{Clusters identified by baseline factors.}
  \label{fig:clusters_baselines}
\end{subfigure}
\begin{subfigure}[t]{0.55\textwidth}
    %\centering
    \includegraphics[width=0.85\linewidth]{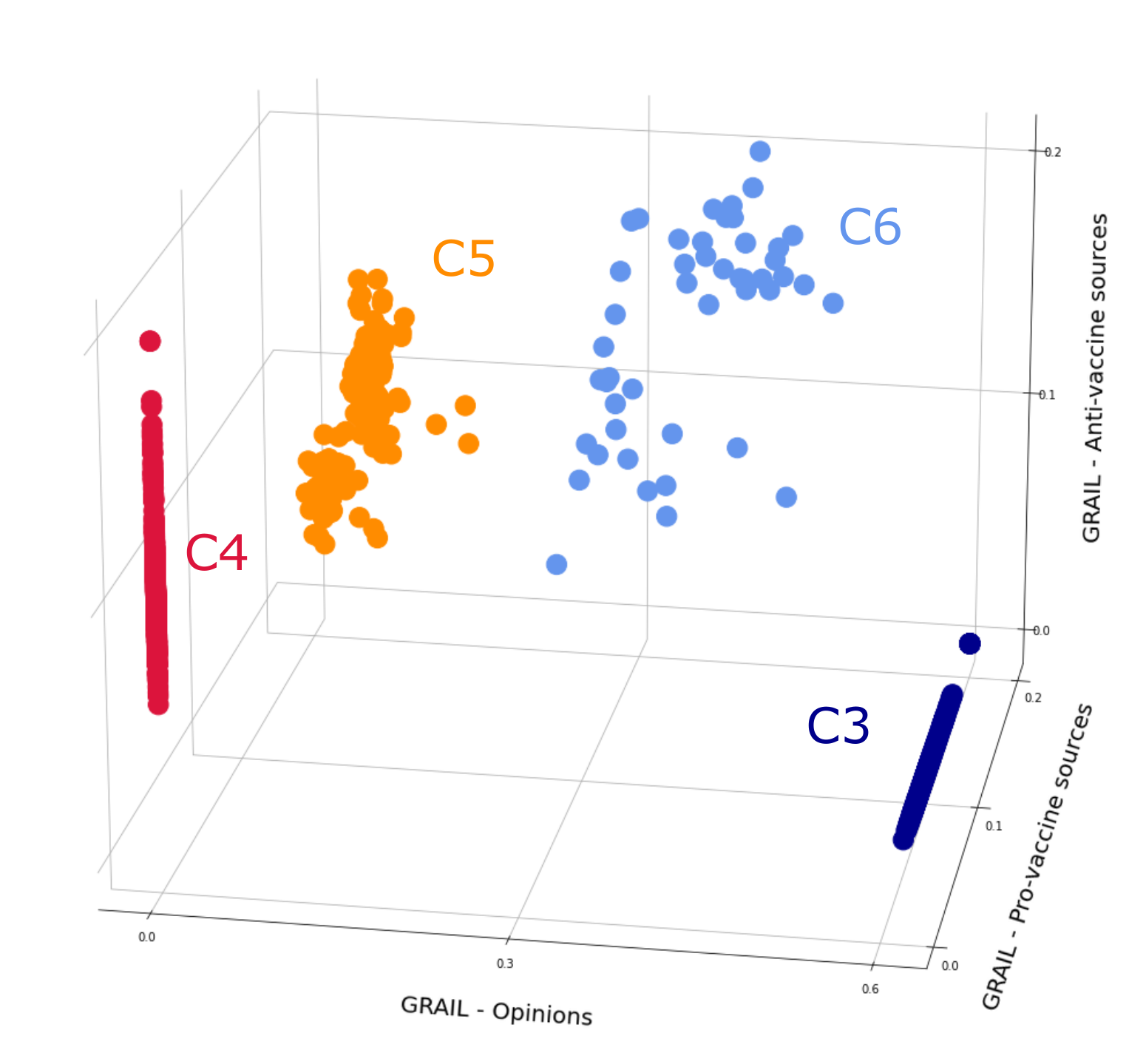}
    \caption{Clusters identified by \modelName factors.}
    \label{fig:clusters_grail_3d}
\end{subfigure}
\caption{Clusters.}
\label{fig:clusters_3D}
\end{figure}

Here, we compare clusters identified through the \modelName factors to those obtained with the baseline factors. These clusters are presented in Figure~\ref{fig:clusters_3D}.

First, looking closely at the 2 clusters obtained with baseline factors (See Figure \ref{fig:clusters_baselines}), we can see that they are only differentiated by the baseline opinion factor, with one cluster corresponding to users who access mostly the anti-vaccine community (noted C1 in Figure \ref{fig:clusters_baselines}), and the other corresponding to users who access mostly the pro-vaccine community (noted C2 in Figure \ref{fig:clusters_baselines}). This was expected given the tight distribution of this factor (See Figure~\ref{fig:distributions}). Due to the heterogeneity of behaviors adopted by users composing $C1$ and $C2$, it is difficult to characterize them more precisely. So, although the two baseline sources factors are evenly distributed, they do not contribute to a clear discrimination of polarization behavior classes: users are only split according to their belonging to either community,  the diversity of sources they interact with does not help to discriminate between them. The polarization behaviors reflected by these clusters are thus difficult to interpret. 

Second, the 4 clusters discriminated based on \modelName factors (See Figure~\ref{fig:clusters_grail_3d}) are differentiated by the three factors: the opinion factor and the two source factors. The polarization behaviors adopted by users in each of these clusters can be more easily characterized:
\begin{enumerate}
    \item \textbf{$C_3$} (452 users) is made up of highly polarized users in the pro-vaccine community. They interact solely within this community, and the diversity of sources with which they interact varies greatly.
    \item \textbf{$C_4$} (360 users) is made up of highly polarized users in the anti-vaccine community. They interact solely within this community, and the diversity of sources with which they interact varies greatly.
    \item \textbf{$C_5$} (141 users) is made up of users who interact mostly in the anti-vaccine community, but have retweeted at least 1 tweet from the pro-vaccine community. In the anti-vaccine community, to which they are closer, users globally interact with a large variety of sources, while they only interact with a limited number of sources in the pro-vaccine community.
    \item \textbf{$C_6$} (47 users) is made up of users who interact in both communities, with a preference for the pro-vaccine community. The distribution of their interactions in each community is more balanced than for users in $C_3$. In the pro-vaccine community, to which they are closer, users globally interact with a variety of sources, while they only interact with a limited number of sources in the anti-vaccine community. 
\end{enumerate}

To sum up, we can say that \modelName factors are not only well-distributed but also that they allow to differentiate and characterize different behaviors (in terms of opinion and source access).

In a nutshell, based on the tuning phase and clustering results, the transformation of entropy-based factors following the polynomial (Equation~(\ref{sigmoid})), as well as their weighting, allow for finely discriminate polarization behaviors and with high clustering performances. Although interesting, we cannot conclude about the reliability of \modelName to evaluate individual polarization. That is why we propose to go further and complete the evaluation of \modelNameNospace, by explaining the differences of behavior adopted in identified clusters based on behavioral factors.

\subsection{Explaining Polarization Scores with Additional Behavioral Indicators}
\label{regression_results}

\begin{table*}[h!]
\centering
  \caption{Optimal combination of indicators for hierarchical regression model for each cluster with coefficients (*p$<$0.05, **p$<$0.01, ***p$<$0.001) and $R^2$ values.}
  \label{tab:regression_results}
  \begin{tabular}{c|ccccc|c}
    \toprule
    Cluster&\multicolumn{5}{|c|}{Ordered combination of indicators}&$R^2$\\
    \hline
    \hline
    $C_3$ &  NRTs & \%vaccine & NPro & & &\textbf{0.81}\\
    &0.01*** & 0.16*** & -0.03*** & & & \\
    \hspace{0.5cm} $\Delta R^2$ & 0.05 & 0.07 & 0.74 & & & \\
    \hline
    $C_4$ & NRTs  & \%vaccine  & NAnti & & & \textbf{0.65}\\
    &-0.01***  & -0.10*** & 0.02*** & & &\\
     \hspace{0.5cm} $\Delta R^2$ & 0.01 & 0.43 & 0.21 &&&\\
     \hline
    $C_5$ & NRTs &  \%vaccine &  \%weeks &NPro& NAnti &\textbf{0.80}\\
    &-0.01*** & -0.24* & -0.03* & 0.07*** & 0,01***&\\
     \hspace{0.5cm} $\Delta R^2$ &  0.01 & 0.16 & 0.01 & 0.6 & 0.02 &\\
     \hline
     $C_6$& NRTs &  \%vaccine &  \%weeks &NPro& NAnti & \textbf{0.61}\\
     &0.01& 0.85 & -0.15 & 0.02*** & -0.12*** & \\
     \hspace{0.5cm} $\Delta R^2$ &  0.01& 0.03 & 0.05 & 0.11 & 0.41 &\\     
  \bottomrule
\end{tabular}
\end{table*}

%\subsubsection{An innovative methodology to characterize polarization behaviors}
To analyze to what extent \modelName provides a reliable evaluation of polarization, we propose a breakthrough evaluation methodology that relies on hierarchical regression (HR). As users adopt well-discriminated behaviors between clusters, we draw the following hypothesis: if the variance of \modelName scores in each cluster can be explained by additional behavioral indicators (\textit{i.e.} not used by \modelNameNospace), it will confirm the ability of \modelName to reflect polarization resulting from different behaviors.  

HR is a statistical model that helps analyze the relationship between a dependent variable and one or more independent variable(s). Concretely, HR provides a final $R^2$ value that represents the proportion of the variability of the dependent variable that can be explained by a combination of independent variables, each allowing to increase this proportion ($\Delta R^2$). It also provides coefficients informing about the relationship between the dependent and independent variables. It gives information about the variation of the dependent variable when the independent varies by one unit.
In this work, \modelName scores correspond to the dependent variable, while additional behavioral indicators (introduced below) are the independent variables. 

Appropriately, the studied dataset allows us to infer information, from which indicators, that are different but complementary to \modelName factors, can be computed. We choose to build indicators, computed from raw data, that represent user engagement, interest, and regularity of interaction in the COVID-19 vaccine debate. For each standard user, we compute the following indicators and apply hierarchical regression:
\begin{itemize}
    \item \textit{NRT} - Number of retweets about the vaccine debate on elite users' tweets;
    \item \textit{\%vaccine} - Proportion of retweets about the vaccine debate in relation to all retweets; 
    \item \textit{\%weeks} - Proportion of active weeks, {\it i.e.} number of weeks in which the user retweeted;
    \item \textit{NPro} - Number of pro-vaccine elite users retweeted;
    \item \textit{NAnti} - Number of anti-vaccine elite users retweeted.
\end{itemize}

 As the size of clusters varies by a factor of 10, we perform a preliminary power test to estimate the required sample size to get robust statistical results and make sure that the results are statistically significant. We set a medium effect size Cohen's $h=0.5$, and the required minimal power of $0.8$, \textit{i.e.} the probability that the hierarchical regression returns an accurate rejection of the null hypothesis. With these parameter values, the minimum required sample size is 34, so results returned by the regression model are robust and well detect statistical differences. Table~\ref{tab:regression_results} presents final $R^2$ values in the rightmost column, as well as the $\Delta R^2$ brought by each indicator and the associated coefficient, whose order and number have been optimized to maximize $R^2$.  

 As a preliminary step, we perform the hierarchical regression on the polarization scores resulting from the combination of baseline factors. The resulting $R^2$ values are satisfactory ($0.35$ for $C_1$ and $0.57$ for $C_2$). However, when we apply it on \modelNameNospace, we notice that $R^2$ values are very high for every behavioral class, ranging between $0.61$ and $0.81$. This confirms that clusters can be better explained through \modelName scores, whatever their size and the polarization behavior they represent. 

 Looking closer at the hierarchical regression results on \modelNameNospace, we notice that the number and type of indicators in the optimal combination differ between clusters: clusters of highly polarized users ($C_3$ and $C_4$) can be explained only with the number of retweets, the proportion of retweets on the COVID-19 vaccine debate, and the number of elite users retweeted in the community of belonging of polarized users. For clusters of users interacting in both communities ($C_5$ and $C_6$), the proportion of active weeks and the number of elite users retweeted in both communities matter. This means that behaviors adopted in each of those clusters can be explained differently, thus behaviors actually differ between identified clusters. In class $C_3$ (highly polarized pro-vaccine users), as we might have expected, \modelName scores are mainly explained by the number of pro-vaccine elites retweeted, with \modelName scores dropping by $0.03$ when one additional pro-vaccine elite user is retweeted (more diverse sources). In the same way, in class $C_4$ (highly polarized users in the anti-vaccine community), \modelName scores increase (getting closer to 0) when the number of anti-vaccine users retweeted increases, as expected. Let us now look at $C_5$ and $C_6$. About $C_5$, which is made up of users who mainly interact with the anti-vaccine community, and to a lesser extent with the pro-vaccine community, numerous indicators contribute to the explanation of the variation of \modelName scores. The interest in the vaccine debate, evaluated through the proportion of interactions on the topic (\textit{\%vaccine}), is important: users interacting a lot on the debate tend to get lower \modelName scores. This means that in $C_5$, the more users interact about the vaccine debate, the more they become polarized towards their majority community. About $C_6$, made up of users with more balanced interactions with both communities and a slight imbalance in favor of the pro-vaccine community, we can notice that the same type of results are obtained: when their proportion of retweets on the debate increases (\textit{\%vaccine}), the polarization scores increases drastically, with an increase of $0.85$. Considering now the frequency of interactions ($\%weeks$), the impact is not the same for users in $C_5$ and $C_6$: with more regular interactions, users of $C_5$ see their polarization score decreasing by $-0.03$, thus translating a higher polarization. On the contrary, when users of $C_6$, whose polarization scores are positive, more frequent interactions imply a decrease of \modelName scores by $0.15$. The effect of the frequency of interactions seems thus to depend on the community to which the user is closer. Besides, results confirm that for $C_5$ and $C_6$, an increase in the number of retweeted pro- or anti-vaccine elite users leads to less extreme \modelName scores.

\section{Conclusion and Perspectives}
\label{CCL}

In this work, we proposed \modelName (\textit{GeneRalized AddItive poLarization}), the first individual and multi-factorial polarization metric. \modelName has three specificities. Each factor is modeled through entropy. The relationship between factors is not linear, it relies on a generalized additive model that uses a polynomial smooth function. \modelName is generalizable and can be adapted to the studied dataset thanks to two main tunable parameters: the shape and the slope of the sigmoid-like curve and the weight of each factor. 

\modelName has been evaluated on a COVID-19 vaccine Twitter dataset, where it is implemented as a 3-factorial metric. \modelName has been compared to a 3-factorial baseline metric, that exploits measures from the literature. Experiments confirm that the use of entropy contributes to a wider distribution of factors.
In addition, \modelName factors, transformed by a polynomial function,  allow the identification of well-separated behavioral clusters. This emphasizes that an adequate combination of factors leads to a reliable assessment of polarization. Besides, through an innovative evaluation framework, based on hierarchical regression, we confirm the relevance of \modelName by finely characterizing each polarization behavioral class. 
These experiments contribute to answer the two RQs raised in the introduction. First, the use of a generalizable individual polarization metric relying on a GAM allows to combine polarization factors accurately \textbf{(RQ1)}. Second, \modelName scores can be well explained in each behavioral class, and thus enables a fine and accurate characterization of adopted behaviors \textbf{(RQ2)}.

In this work, we have focused on polarization in the specific context of social media, by exploiting a Twitter dataset. Nevertheless, \modelName's components make it a good candidate for assessing polarization in a broader context. Especially in a wider perspective of access to information, notably via recommendation systems. The multiple polarization factors processed by \modelName can be adapted and determined according to the context of the study and associated data. Going further, as the polarization is highly dynamic, we also wonder how  \modelName could be exploited to assess the temporal process of polarization. Experiments on this temporal dimension are currently being carried out. 

In the meantime, the version of \modelName proposed in this paper allows a finer understanding of polarization behaviors on social media and brings new perspectives to take charge of online polarization. Among other things, the need for personalized depolarization strategies has been put forward in the literature: a joint approach is no longer sufficient~\cite{stray2021designingrs, bernstein2020diversity, treuillier2022being}. In this respect, the unprecedented characterization of polarization behaviors based on \modelName is of great interest. It will contribute to our future works related to the development of depolarization algorithms, used by recommender systems. Such systems can propose recommendations tailored to each user's behavior, with the goal of controlling the effect of their depolarizing capacity.

% \begin{acks}

% \end{acks}

\paragraph{Aknowledgments}
This research was supported by BOOM (Modeling and Opening Opinion Bubbles) (ANR-20-CE23-0024).  

\bibliographystyle{unsrt}
\bibliography{refs}

\begin{thebibliography}{10}

\bibitem{benkler2018network}
Yochai Benkler, Robert Faris, and Hal Roberts.
\newblock {\em Network propaganda: Manipulation, disinformation, and
  radicalization in American politics}.
\newblock Oxford University Press, 2018.

\bibitem{choi2016concept}
Moonsun Choi.
\newblock A concept analysis of digital citizenship for democratic citizenship
  education in the internet age.
\newblock {\em Theory \& research in social education}, 44(4):565--607, 2016.

\bibitem{tucker2018social}
Joshua~A Tucker, Andrew Guess, Pablo Barber{\'a}, Cristian Vaccari, Alexandra
  Siegel, Sergey Sanovich, Denis Stukal, and Brendan Nyhan.
\newblock Social media, political polarization, and political disinformation: A
  review of the scientific literature.
\newblock {\em Political polarization, and political disinformation: a review
  of the scientific literature (March 19, 2018)}, 2018.

\bibitem{kubin2021role}
Emily Kubin and Christian von Sikorski.
\newblock The role of (social) media in political polarization: a systematic
  review.
\newblock {\em Annals of the International Communication Association},
  45(3):188--206, 2021.

\bibitem{stray2021designingrs}
Jonathan Stray.
\newblock Designing recommender systems to depolarize.
\newblock {\em First Monday}, 27, 2021.

\bibitem{prior2013media}
Markus Prior.
\newblock Media and political polarization.
\newblock {\em Annual Review of Political Science}, 16:101--127, 2013.

\bibitem{helberger2018diversity}
Natali Helberger, Kari Karppinen, and Lucia D’Acunto.
\newblock Exposure diversity as a design principle for recommender systems.
\newblock {\em Information, Communication \& Society}, 21(2):191--207, 2018.

\bibitem{lunardi2020filter}
Gabriel~Machado Lunardi, Guilherme~Medeiros Machado, Vinicius Maran, and José
  Palazzo~M. {de Oliveira}.
\newblock A metric for filter bubble measurement in recommender algorithms
  considering the news domain.
\newblock {\em Applied Soft Computing}, 97:106771, 2020.

\bibitem{heitz2022benefits}
Lucien Heitz, Juliane~A Lischka, Alena Birrer, Bibek Paudel, Suzanne Tolmeijer,
  Laura Laugwitz, and Abraham Bernstein.
\newblock Benefits of diverse news recommendations for democracy: A user study.
\newblock {\em Digital Journalism}, pages 1--21, 2022.

\bibitem{bernstein2020diversity}
Abraham Bernstein, Claes De~Vreese, Natali Helberger, Wolfgang Schulz,
  Katharina Zweig, Christian Baden, Michael~A Beam, Marc~P Hauer, Lucien Heitz,
  Pascal J{\"u}rgens, et~al.
\newblock Diversity in news recommendations.
\newblock {\em arXiv preprint arXiv:2005.09495}, 2020.

\bibitem{treuillier2022being}
Celina Treuillier, Sylvain Castagnos, Evan Dufraisse, and Armelle Brun.
\newblock Being diverse is not enough: Rethinking diversity evaluation to meet
  challenges of news recommender systems.
\newblock In {\em Fairness in User Modeling, Adaptation and Personalization
  (FairUMAP 2022)}, 2022.

\bibitem{garimella2018quantifying}
Kiran Garimella, Gianmarco De~Francisci Morales, Aristides Gionis, and Michael
  Mathioudakis.
\newblock Quantifying controversy on social media.
\newblock {\em ACM Transactions on Social Computing}, 1(1):1--27, 2018.

\bibitem{guerra2013measure}
Pedro Guerra, Wagner Meira~Jr, Claire Cardie, and Robert Kleinberg.
\newblock A measure of polarization on social media networks based on community
  boundaries.
\newblock In {\em Proceedings of the international AAAI conference on web and
  social media}, volume~7, pages 215--224, 2013.

\bibitem{morales2015measuring}
Alfredo~Jose Morales, Javier Borondo, Juan~Carlos Losada, and Rosa~M Benito.
\newblock Measuring political polarization: Twitter shows the two sides of
  venezuela.
\newblock {\em Chaos: An Interdisciplinary Journal of Nonlinear Science},
  25(3):033114, 2015.

\bibitem{becatti2019extracting}
Carolina Becatti, Guido Caldarelli, Renaud Lambiotte, and Fabio Saracco.
\newblock Extracting significant signal of news consumption from social
  networks: the case of twitter in italian political elections.
\newblock {\em Palgrave Communications}, 5(1):1--16, 2019.

\bibitem{cicchini2022news}
Tomas Cicchini, Sofia~Morena Del~Pozo, Enzo Tagliazucchi, and Pablo Balenzuela.
\newblock News sharing on twitter reveals emergent fragmentation of media
  agenda and persistent polarization.
\newblock {\em EPJ Data Science}, 11(1):48, 2022.

\bibitem{geschke2019triple}
Daniel Geschke, Jan Lorenz, and Peter Holtz.
\newblock The triple-filter bubble: Using agent-based modelling to test a
  meta-theoretical framework for the emergence of filter bubbles and echo
  chambers.
\newblock {\em British Journal of Social Psychology}, 58(1):129--149, 2019.

\bibitem{jost2022cognitive}
John~T Jost, Delia~S Baldassarri, and James~N Druckman.
\newblock Cognitive--motivational mechanisms of political polarization in
  social-communicative contexts.
\newblock {\em Nature Reviews Psychology}, 1(10):560--576, 2022.

\bibitem{valensise2023drivers}
Carlo~M Valensise, Matteo Cinelli, and Walter Quattrociocchi.
\newblock The drivers of online polarization: fitting models to data.
\newblock {\em Information Sciences}, page 119152, 2023.

\bibitem{phillips2023organizational}
Samantha~C Phillips and Kathleen~M Carley.
\newblock An organizational form framework to measure and interpret online
  polarization.
\newblock {\em Information, Communication \& Society}, pages 1--33, 2023.

\bibitem{treuillier23}
Celina Treuillier, Sylvain Castagnos, and Armelle Brun.
\newblock {A Multi-Factorial Analysis of Polarization on Social Media}.
\newblock In {\em {UMAP'23}}, Limassol, Cyprus, June 2023.

\bibitem{sirbu2017opinion}
Alina S{\^\i}rbu, Vittorio Loreto, Vito~DP Servedio, and Francesca Tria.
\newblock Opinion dynamics: models, extensions and external effects.
\newblock In {\em Participatory sensing, opinions and collective awareness},
  pages 363--401. Springer, 2017.

\bibitem{baumann2020modeling}
Fabian Baumann, Philipp Lorenz-Spreen, Igor~M Sokolov, and Michele Starnini.
\newblock Modeling echo chambers and polarization dynamics in social networks.
\newblock {\em Physical Review Letters}, 124(4):048301, 2020.

\bibitem{chen2019modeling}
Tinggui Chen, Qianqian Li, Jianjun Yang, Guodong Cong, and Gongfa Li.
\newblock Modeling of the public opinion polarization process with the
  considerations of individual heterogeneity and dynamic conformity.
\newblock {\em Mathematics}, 7(10):917, 2019.

\bibitem{garimella2021political}
Kiran Garimella, Tim Smith, Rebecca Weiss, and Robert West.
\newblock Political polarization in online news consumption.
\newblock In {\em Proceedings of the International AAAI Conference on Web and
  Social Media}, volume~15, pages 152--162, 2021.

\bibitem{michiels2022filter}
Lien Michiels, Jens Leysen, Annelien Smets, and Bart Goethals.
\newblock What are filter bubbles really? a review of the conceptual and
  empirical work.
\newblock In {\em Adjunct Proceedings of the 30th ACM Conference on User
  Modeling, Adaptation and Personalization}, pages 274--279, 2022.

\bibitem{pariser2011filter}
Eli Pariser.
\newblock {\em The filter bubble: How the new personalized web is changing what
  we read and how we think}.
\newblock Penguin, 2011.

\bibitem{bail2018opposing}
Christopher~A. Bail, Lisa~P. Argyle, Taylor~W. Brown, John~P. Bumpus, Haohan
  Chen, M.~B.~Fallin Hunzaker, Jaemin Lee, Marcus Mann, Friedolin Merhout, and
  Alexander Volfovsky.
\newblock Exposure to opposing views on social media can increase political
  polarization.
\newblock {\em Proceedings of the National Academy of Sciences},
  115(37):9216--9221, 2018.

\bibitem{wu2018personalizing}
Wen Wu, Li~Chen, and Yu~Zhao.
\newblock Personalizing recommendation diversity based on user personality.
\newblock {\em User Modeling and User-Adapted Interaction}, 28(3):237--276,
  2018.

\bibitem{vanbavel2021media}
Jay~J. Van~Bavel, Steve Rathje, Elizabeth Harris, Claire Robertson, and Anni
  Sternisko.
\newblock How social media shapes polarization.
\newblock {\em Trends in Cognitive Sciences}, 25(11):913--916, 2021.

\bibitem{conover2011political}
Michael Conover, Jacob Ratkiewicz, Matthew Francisco, Bruno Gon{\c{c}}alves,
  Filippo Menczer, and Alessandro Flammini.
\newblock Political polarization on twitter.
\newblock In {\em Proceedings of the international aaai conference on web and
  social media}, volume~5, pages 89--96, 2011.

\bibitem{newman2006modularity}
Mark~EJ Newman.
\newblock Modularity and community structure in networks.
\newblock {\em Proceedings of the national academy of sciences},
  103(23):8577--8582, 2006.

\bibitem{schmidt2018polarization}
Ana~Luc{\'\i}a Schmidt, Fabiana Zollo, Antonio Scala, Cornelia Betsch, and
  Walter Quattrociocchi.
\newblock Polarization of the vaccination debate on facebook.
\newblock {\em Vaccine}, 36(25):3606--3612, 2018.

\bibitem{sunstein1999law}
Cass~R Sunstein.
\newblock The law of group polarization.
\newblock {\em University of Chicago Law School, John M. Olin Law \& Economics
  Working Paper}, (91), 1999.

\bibitem{shannon1948mathematical}
Claude~Elwood Shannon.
\newblock A mathematical theory of communication.
\newblock {\em The Bell system technical journal}, 27(3):379--423, 1948.

\bibitem{hastie2017generalized}
Trevor~J Hastie.
\newblock Generalized additive models.
\newblock In {\em Statistical models in S}, pages 249--307. Routledge, 2017.

\bibitem{primario2017measuring}
Simonetta Primario, Dario Borrelli, Luca Iandoli, Giuseppe Zollo, and Carlo
  Lipizzi.
\newblock Measuring polarization in twitter enabled in online political
  conversation: The case of 2016 us presidential election.
\newblock In {\em 2017 IEEE international conference on information reuse and
  integration (IRI)}, pages 607--613. IEEE, 2017.

\bibitem{clauset2004finding}
Aaron Clauset, Mark~EJ Newman, and Cristopher Moore.
\newblock Finding community structure in very large networks.
\newblock {\em Physical review E}, 70(6):066111, 2004.

\bibitem{likas_global_2003}
Aristidis Likas, Nikos Vlassis, and Jakob J.~Verbeek.
\newblock The global k-means clustering algorithm.
\newblock {\em Pattern Recognition}, pages 451--461, 2003.

\bibitem{davies1979cluster}
David~L Davies and Donald~W Bouldin.
\newblock A cluster separation measure.
\newblock {\em IEEE transactions on pattern analysis and machine intelligence},
  (2):224--227, 1979.

\bibitem{rousseeuw_silhouettes_1987}
Peter~J. Rousseeuw.
\newblock Silhouettes: A graphical aid to the interpretation and validation of
  cluster analysis.
\newblock {\em Journal of computational and applied mathematics}, pages 53--65,
  1987.

\end{thebibliography}

\end{document}